# Investigation of Vertical Spiral Resonators for Low Frequency Metamaterial Design

## J. Zhu, T. Hao, C. J. Stevens and D. J. Edwards

Department of Engineering Science, University of Oxford
Parks Road, Oxford, OX1 3PJ, UK
email: jiwen.zhu@eng.ox.ac.uk

**Abstract**

This paper thoroughly explores the characteristics of vertical spiral resonators (VSR). They exhibit relatively high Q factors and sizes around a few percent of the free space wavelength, which make them ideal candidates for assembling metamaterial devices. A quasistatic model of VSR is obtained from simple analytical expressions, and the effects of certain geometrical parameters on the resonant frequency are investigated.

## 1. Introduction

Electrically small resonant particles are key components for metamaterial design so as to realise negative equivalent material parameters, and the miniaturisation of these particles is crucial if the metamaterial has to be viewed as a continuous medium. The split ring resonator (SRR) was originally adopted for the realisation of a negative magnetic permeability medium [1] and subsequently a left-handed medium [2]. Later other resonators have been introduced including broadside coupled SRR (BC-SRR) [3], spiral resonator [4] and vertical spiral resonator (VSR) [5], which have electrically smaller sizes and more isotropic behaviour.

The VSR has been demonstrated to exhibit a relatively high Q factor compared to other common resonator topologies [5]. In this paper the quasistatic LC model for BC-SRR [3] is applied to investigate the resonance of VSR, but the equivalent inductance is obtained via a simple analytical formula. The results of our proposed model are compared with that of the original model and the EM full wave simulation. Furthermore, the effects of various geometrical parameters of VSR on its resonant frequency have been investigated which provide some useful conclusions for the resonator designs.

## 2. Model for the Vertical Spiral Resonator

The four different resonator structures involved in this paper are represented in Fig.1. The SRR and spiral resonator are of single side configurations, while the BC-SRR and VSR are formed by two metallic rings with same dimensions printed on both sides of dielectric slab. A conducting via through the substrate connects the two rings of VSR together. This vertical configuration is aimed to achieve further miniaturisation of element size since the distributed capacitance between the rings can be significantly enhanced by broadside coupling as compared to single side edge coupling.

The quasistatic LC model [3] is exploited for all of the above topologies, which assumes that the sum of the currents on each ring is uniform for a given value of the angular polar coordinate $\phi$. Under this assumption, the equivalent inductances are roughly the same for all the four structures with similar dimensions, and can be approximated by the inductance of a single equivalent ring with the same average radius and a width equal to the width of each original ring. However, that inductance in [3] is obtained via an infinite integral which requires numerical solution and therefore introduces additional complexity to the problem.

Instead of performing numerical integration, we have adopted an accurate analytical formula [6] which was originally developed for evaluating the inductance of planar spiral inductors and bears a typical error range of around 1-2%. The expression for the inductance of a single ring is determined as

$$L = \mu_0 r_0 \left[\ln(4.92 r_0/W) + 0.2(W/2r_0)^2\right] \quad (1)$$

where the geometrical parameters $r_0$ and $W$ are as shown in Fig.1(d).

In [4], it has been demonstrated that under quasistatic approximation, the equivalent circuit capacitance of a SRR is the series connection of the capacitance of its upper and lower halves, while that of a spiral resonator is the parallel connection of the two halves. Therefore the global capacitance of a spiral resonator is roughly four times that of a SRR with the same geometrical parameters. As can be easily appreciated, the quasistatic field distribution of BC-SRR and VSR are similar with that of SRR and spiral resonator respectively. Therefore we can deduce that the global equivalent capacitance of a VSR is also four times that of a BC-SRR with the same dimensions, and the resonant frequency can thus be roughly reduced by half. The capacitance is obtained following the same procedure as in [3], i.e., the total capacitance of a VSR is

$$C = \pi r_0 \sqrt{\varepsilon_e}/(c_0 Z_0) \quad (2)$$

where $c_0$ is the speed of EM wave in free space, $Z_0$ is the characteristic impedance of a microstrip line with the metallic strip width $W$ and substrate thickness $t/2$. $\varepsilon_e$ is the effective dielectric constant of the above microstrip line. The curvature effects are neglected and error would thus be introduced, which would be significant if the strip width is comparable with the ring radius.

After obtaining the equivalent circuit parameters, the resonant frequency is calculated as $1/2\pi\sqrt{LC}$.

## 3. Influence of geometrical parameters on the resonant frequency of VSR

In this section, the resonant frequencies of VSRs with different geometrical parameters were calculated using the developed model, the model for BC-SRR [3] and full-wave EM simulation software Micro-Stripes® [7]. The results are compared in order to verify the accuracy of the proposed model. Besides, the effects of various geometrical parameters on the resonant frequency are concluded.

The relative dielectric constant of the substrate is $\varepsilon_r = 3.1$ and the metallisation thickness is 18 $\mu$m, which corresponds to the available polyimide sheet. Initially the parameters were taken to be $r_0 = 3.5$mm, $W = 1$mm, $t = 50$ $\mu$m. By keeping others constant, we only changed one parameter at a time.

First, the substrate thickness $t$ took different values with the calculated frequencies shown in Fig.2(a). As can been seen, increasing the substrate thickness will increase the resonant frequency. Secondly, the strip width $W$ was varied with the calculation results shown in Fig.2(b). Smaller strip width will result in a higher resonant frequency. Finally, the mean radius of the ring $r_0$ was varied and the corresponding frequencies are depicted in Fig.2(c). Larger rings would have smaller resonant frequencies. Among all the situations, the results obtained from different models agree well with each other with a typical discrepancy of within 2%, which tends to be bigger for rings with larger diameters. In addition, all of the above VSRs have diameters of only several percent of the free space wavelengths, thus a significant level of miniaturisation have been achieved by exploiting the vertical spiral configuration with very thin dielectric layers.

## 4. Conclusion

In this paper, a quasistatic model for the VSR based on analytical formulae is developed, whose accuracy is of the same level as both the currently available analytical model and full-wave EM simulation. The resonant frequency of VSR can be greatly reduced by decreasing the substrate thickness, increasing the metallic strip width and using rings with larger diameters. By utilising very thin substrates with thickness of tens of microns, the sizes of VSRs can be reduced to only a few percent of the free space wavelength, which makes them useful artificial atoms to better approach continuous artificial media. Furthermore, such thin substrates are quite flexible, which could realise many novel potential applications such as conformal components.

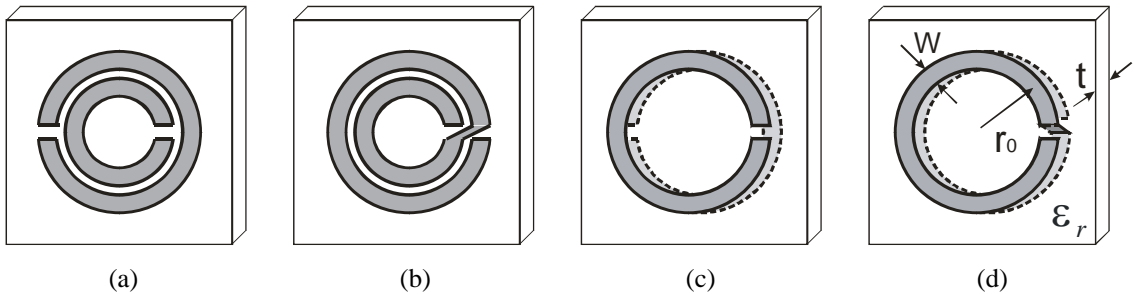

(a)     (b)     (c)     (d)

Fig. 1: Four resonator structures. (a) Split ring resonator (SRR). (b) Spiral resonator. (c) Broadside coupled SRR (BC-SRR). (d) Vertical spiral resonator (VSR).

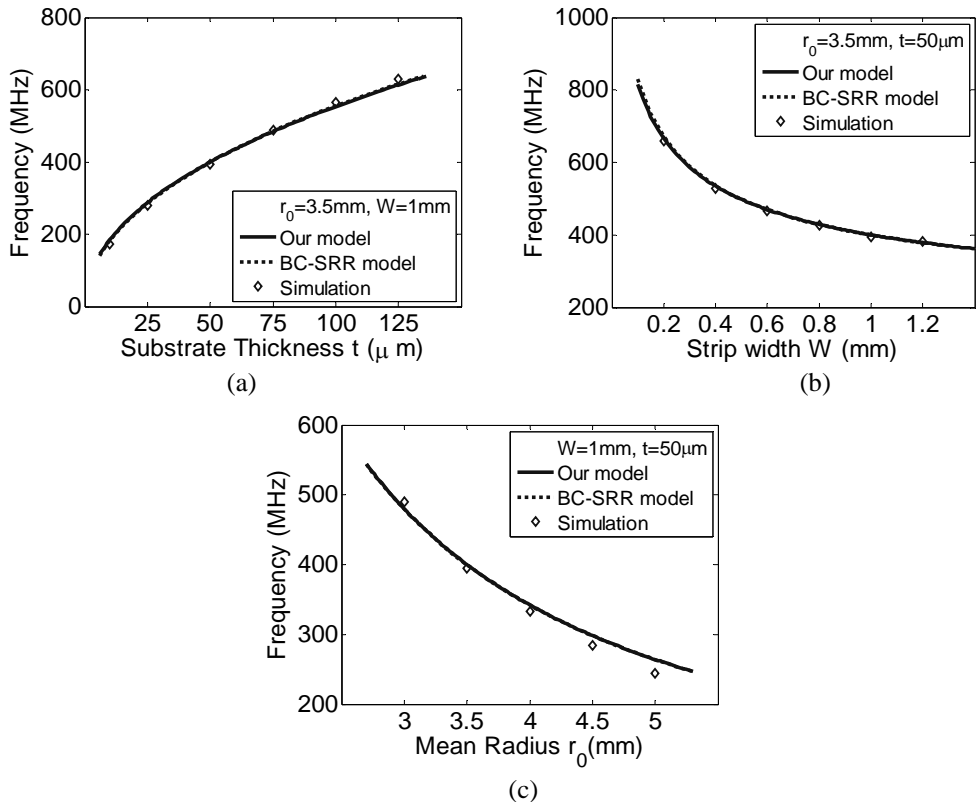

(a)     (b)

(c)

Fig. 2: The effects of geometrical parameters on the resonant frequency of VSR. (a) Substrate thickness. (b) Strip width. (c) Mean radius.